\documentclass[english,aps,manuscript]{revtex4}
\usepackage[T1]{fontenc}
\usepackage[latin1]{inputenc}

\makeatletter

\makeatletter

\newcommand{\bee}{\begin{equation}}
\newcommand{\ee}{\end{equation}}
\newcommand{\beea}{\begin{eqnarray}}
\newcommand{\eea}{\end{eqnarray}}

\makeatother

\usepackage{babel}
\makeatother
\begin{document}

\preprint{COLO-HEP-525, NSF-KITP-06-147}

\maketitle
\textbf{\Large String Phenomenology and the Cosmological Constant }{\Large \par}

\begin{center}\vspace{0.3cm}
\par\end{center}

\begin{center}{\large S. P. de Alwis$^{\dagger}$ }\par\end{center}{\large \par}

\begin{center}\vspace{0.3cm}
\par\end{center}

\begin{center}Physics Department, University of Colorado, \\
 Boulder, CO 80309 USA\par\end{center}

\begin{center}\vspace{0.3cm}
\par\end{center}

\begin{center}\textbf{Abstract}\par\end{center}

\begin{center}\vspace{0.3cm}
\par\end{center}

\begin{center}It is argued that classical string solutions should
not be fine tuned to have a positive cosmological constant (CC) at
the observed size, since even the quantum corrections from standard
model effects will completely negate any classical string theory solution
with such a CC. In fact it is even possible that there is no need
at all for any ad hoc uplifting term in the potential since these
quantum effects may well take care of this. Correspondingly any calculation
of the parameters of the MSSM has to be rethought to take into account
the evolution of the CC. This considerably complicates the issue since
the initial conditions for RG evolution of these parameters are determined
by the final condition on the CC! The Anthropic Principle is of no
help in addressing these issues. \par\end{center}

\vspace{0.3cm}

PACS numbers: 11.25. -w, 98.80.-k; 

\vfill{}

$^{\dagger}$ {\small e-mail: dealwis@pizero.colorado.edu}{\small \par}

\eject

\section{Introduction}

Much effort has been expended on trying to find constructions which
yield solutions with a small positive cosmological constant (CC) in
the context of recent work on flux compactifications of string theory.
The motivation stems from the observation of a small positive CC.
These classical string solutions are a good description at some high
scale somewhat less than the string or Kaluza-Klein (KK) scale. But
the observed cosmological constant is measured at cosmological scales,
and clearly one needs to incorporate quantum effects before one can
compare the CC of the effective theory coming from the string theory
model, with the observed cosmological constant.

In this respect it is necessary to clear up some apparent confusion
in the literature concerning the argument of Bousso and Polchinski
(BP) and its relation to the concrete models of classical string solutions
in which all the moduli are stabilized. The argument of BP proceeds
from the following assertion: the measured cosmological constant $\Lambda$
can be written, in the context of string theory, as the sum of two
contributions\begin{equation}
\Lambda=\Lambda_{0}+\sum_{i}n_{i}^{2}q_{i}^{2}.\label{BT}\end{equation}
Here the second term is the contribution of internal fluxes and the
$q_{i}$ are a set of charges and the $n_{i}$ are a set of integers
characterizing the set of fluxes through the various cycles of the
internal space. $\Lambda_{0}$ (which is assumed to be negative) is
the sum of all contributions except those coming from the fluxes -
\textit{in particular it is supposed to include all quantum corrections
to the cosmological constant}. However the arguments of BP were not
made in a context where the moduli were stabilized, and in any concrete
context where this has been achieved, the corresponding formula including
quantum corrections is much more complicated than what is indicated
by (\ref{BT}).

Consider the case of type IIB compactifications. The work of many
authors (see \cite{Grana:2005jc} for a recent review) led to the
realization that the complex structure moduli and the dilaton could
be stabilized by turning on internal fluxes. Then it was pointed out
in \cite{Kachru:2003aw} (KKLT) that with the inclusion of non-perturbative
terms the Kaehler moduli could also be stabilized. However this calculation
resulted in a minimum which was negative. KKLT then decided to raise
this minimum by adding an uplifting term, originating from Dbar branes,
in order to get a model with a positive cosmological constant. Subsequently
many authors considered other mechanisms for uplifting AdS minima
to dS minima %
\footnote{For a list of references see \cite{Grana:2005jc}.%
}.

However there clearly is no need to insist on getting a positive CC
at the observed value from a classical string theory calculation,
which can only give the initial conditions for an RG evolution down
to the far infra-red. KKLT type calculations are essentially done
in classical string theory (or rather its low energy supergravity
(SUGRA) approximation) with some instanton corrections. Even though
some attempts have been made to include $\alpha'$ corrections, the
question of including the corrections coming from integrating down
from the string/KK scale has not been seriously addressed. Instead
what many authors have done is to start with a KKLT type model (or
one of its variants) with the cosmological constant adjusted to the
observed value. Then the minimal supersymmetric standard model (MSSM)
parameters (Yukawa couplings, soft masses) are computed. The latter
are then used as the initial conditions of an RG evolution down to
the standard model scale.

What is ignored in these calculations is the fact that in going from
the KK scale down to the TeV scale the cosmological constant also
evolves - in fact (generically) it acquires corrections which are
quadratic in the cutoff scale (quartic corrections are absent in a
theory with equal numbers of bosonic and fermionic degrees of freedom).
In very special situations though it may be possible to make sure
that these quadratic corrections are absent (see for example (\cite{Ferrara:1994kg})
and then the corrections are of order the supermultiplet splitting
(TeV?) scale. Thus if one started with a model with a tiny positive
CC at the observed value, one ends up with a cosmological constant
which is at least at the TeV scale! Clearly such a calculation makes
no sense.

It was in fact pointed out in \cite{Brustein:2004xn} that at the
classical string theory level what one needs to ensure is that the
cosmological constant is at or below the TeV scale. One could very
well have started with a negative CC at the classical level, and (provided
that this minimum breaks supersymmetry (SUSY)) one could envisage
(depending on the model) the subsequent evolution to produce positive
contributions that could lift it up. Alternatively it is also possible
to start with a classically SUSY (AdS) minimum of the moduli potential
and then consider (provided this minimum was tuned to be at the TeV
scale) an uplift coming from dynamics of some hidden matter sector
as in \cite{Intriligator:2006dd}. One would still be left with the
usual (broken SUSY) fine-tuning problem of the CC at the level of
(at least) 1 part in $10^{60}$. In any event it is important to realize
that the simple cancellation posited in equation (\ref{BT}) is no
longer valid. The quantum corrections are themselves dependent on
the fluxes since the masses of the moduli and all matter couplings
and masses and hence in particular the splitting within supermultiplets
(whatever the mechanism of SUSY breaking) are dependent on them.

There is an additional problem that to the author's knowledge has
not been discussed in the literature. Suppose that a string theory
construction of the standard model (or more likely the MSSM) is found.
There is likely to be a large degeneracy in that the same MSSM states
are likely to emerge from many different choices of internal manifolds
and flux configurations. Let us call the set of values for any such
choice $\beta$. The four dimensional moduli potential will depend
on this choice i.e. $V=V(\Phi,\beta)$, where $\Phi$ are the moduli
and the MSSM fields. The MSSM parameters i.e. the Yukawa couplings
and the soft masses and the CC will therefore depend on $\beta$.
Now in order to have any predictive power at all, one should be able
to find at most a few values of $\beta$ which give the observed CC
and SM couplings, from which the MSSM parameters could then be computed.
One might hope that the probability distribution on the landscape
is highly peaked at such values of $\beta$. Then one could hope to
compute the MSSM parameters and compare with experiments in the near
future (assuming low energy SUSY is observed). But therein lies the
problem. The allowed set of values $\beta$ are not directly determined
by the experimental value of the cosmological constant. In order to
determine them one needs to evolve the observed value back to the
ultra violet scale at which the classical CC is relevant. But this
evolution depends on the states that are integrated out in between
these two scales and which are in turn dependent on $\beta$. In other
words what we have is a highly non-linear coupled problem.

In the following we elaborate on this.

\section{One loop effective potential\label{sec:One-loop-effective}}

Using the Coleman-Weinberg \cite{Coleman:1973jx} formula the effective
potential upto one loop corrections is \begin{equation}
V=V_{c}+\frac{1}{32\pi^{2}}{\rm Str}M^{2}\Lambda^{2}+\frac{1}{64\pi^{2}}{\rm Str}M^{4}\ln\frac{M^{2}}{\Lambda^{2}}+\ldots.\label{pot1lp}\end{equation}
 Here $V_{c}$ is the classical potential, the ellipses represent
cutoff ($\Lambda$) independent terms, and\begin{equation}
{\rm Str}M^{n}\equiv\sum_{i}(-1)^{2J_{i}+1}m_{i}^{n}(\Phi),\label{suptrace}\end{equation}
where $m_{i}$ is the (field dependent) mass of a particle with spin
$J_{i}$. Note also that the quartic divergence is zero because its
coefficient ${\rm Str}M^{0}=0$ in a theory with equal numbers of
bosonic and fermionic degrees of freedom. Now if SUSY is softly broken
then ${\rm Str}M^{2}$ should be field independent. However there
could still be a contribution to the cosmological constant of $O(m_{3/2}^{2}\Lambda^{2})$
since the SUSY breaking supertrace is of the order of the gravitino
mass %
\footnote{Note that this is true irrespective of the mode of supersymmetry breaking.
It just follows from formula \ref{strm^2} below.%
}. In a string theory calculation of these one loop effects one would
expect a similar formula to hold in the low energy limit with the
cutoff replaced by the string scale %
\footnote{There is a complication here that we will ignore. This comes from
the fact that the Planck scale to string scale ratio is actually a
modulus.%
}. 

The classical potential is ( in Planck units $M_{p}^{2}\equiv\frac{1}{8\pi G_{N}}=1$)
\begin{eqnarray}
V_{c} & = & V_{F}+V_{D}.\nonumber \\
V_{F} & = & e^{G}(G_{i}G_{\bar{j}}G^{i\bar{j}}-3),\label{potclassical}\\
V_{D} & = & (f_{R}^{-1})^{ab}k_{a}^{i}k_{b}^{j}G_{i}G_{j},\nonumber \end{eqnarray}
 Here $G=K(\Phi,\bar{\Phi})+\ln|W(\Phi)|^{2}$ is the Kaehler invariant
combination of the Kaehler potential $K$ and the superpotential $W(\Phi)$.
Also $G_{i}=\partial_{i}G$, $G_{i\bar{j}}=\partial_{i}\partial_{\bar{j}}G$
is the Kaehler metric, and $k_{a}$ is a Killing vector corresponding
to some gauge symmetry generator labelled by the index $a$. Of course
this general form of the potential is expected to be valid quantum
mechanically as well, but the point is that the expression for the
Kaehler potential $K$ that one starts with is taken from a classical
string theory calculation.

First assume that the classical potential has a minimum at $V_{F}=V_{D}=0$.
Now in (\ref{pot1lp}) the coefficient of the quadratic term in the
cutoff is given by\begin{equation}
{\rm Str}M^{2}(\Phi,\bar{\Phi})=2Q(\Phi,\bar{\Phi})m_{3/2}^{2}(\Phi,\bar{\Phi}).\label{strm^2}\end{equation}
 Here $m_{3/2}^{2}=e^{K}|W|^{2}$, and assuming for simplicity that
SUSY is broken in some modulus direction $T$, \begin{equation}
Q(\Phi,\bar{\Phi})=N_{tot}-1-G^{T}H_{T\bar{T}}G^{\bar{T}}.\label{Q}\end{equation}
 Now we have from (\ref{potclassical}) and the assumption that the
minimum has zero CC, $|G_{T}|\sim O(\sqrt{3})$. $H_{T\bar{T}}$ is
generically of $O(1)$ (see eqn (\ref{Hijbar}) for a definition).
Then we see that for large values of $N_{tot}$ (the total number
of degrees of freedom in the effective theory) the quadratic contribution
to the potential is positive. Of course it is not clear that we will
have an extremum, leave alone a minimum, but clearly the possibility
of getting a dS minimum exists!

Although ${\rm Str}M^{2}$ needs to be independent of the MSSM fields
in order to have soft supersymmetry breaking, it can still be a function
of the moduli fields. In any case generically it is non-zero and gives
an $O(m_{3/2}^{2}\Lambda^{2})$ contribution to the cosmological constant.
Under some special circumstances \cite{Ferrara:1994kg} this could
be zero, in which case the leading perturbative contribution would
be of $O(\Delta m^{4}\sim(1TeV)^{4})$ where the last estimate follows
from the observational bound on SUSY partners to the SM fields. In
other words even under the most favorable circumstances the classical
CC needs to be fine tuned to sixty decimal places. Note also that
the above formulae are independent of the mode of transmission of
supersymmetry breaking from a hidden sector to the visible sector.
The general conclusion is valid for any mechanism for mediating supersymmetry
breaking (including low energy mechanisms such as gauge mediation)
since it depends only on the Coleman-Weinberg formula (\ref{pot1lp}).

Actually the above formulae are valid only if the classical starting
point is in flat space. To really be consistent one should recalculate
these perturbative corrections around an arbitrary (curved) background.
Below we will briefly review some aspects of such calculations and
discuss some of the additional complications involved.

\section{Curved space calculations\label{sec:Curved}}

The superpotential $W$ cannot get perturbative corrections but the
Kaehler potential can. Now in global supersymmetry the order parameter
for supersymmetry breaking is $F_{i}=\partial_{i}W$ (i.e. the derivative
of $W$ with respect to any chiral scalar $\Phi^{i}$) at the minimum,
so that if the theory has no supersymmetry breaking minima classically,
then perturbative corrections will not generate one (ignore D-terms
for the moment). In local supersymmetry however the situation is more
subtle. The (F-term) potential for chiral scalars can be rewritten
as \begin{equation}
V=e^{K}(F_{i}\bar{F}_{\bar{j}}K^{i\bar{j}}-3|W|^{2})\label{SUGRAPOT}\end{equation}

Here $K=K(\Phi_{i,}\Phi_{\bar{j}}),\, W=W(\Phi_{i}),\textrm{ F}_{i}=\partial_{i}W+\partial_{i}KW$
and $K_{i\bar{j}}=\partial_{i}\partial_{\bar{j}}K$ is the Kaehler
metric. The order parameter $F_{i}$ now involves both $W$ and $K$,
but since the latter is renormalized in perturbation theory, in general
one would expect that the condition for supersymmetry $F_{i}=0,\,\forall i$,
could be affected in perturbation theory. Around a Poincare invariant
vacuum (i.e one with zero cosmological constant (CC)) however this
cannot happen. The reason is that for a flat space supersymmetry the
two terms in $V$ must cancel at the minimum which in effect means
that the condition for supersymmetry is $\partial_{i}W=W=0$ at the
minimum. Then since $W$ is not renormalized in perturbation theory,
one cannot break supersymmetry. The other (and in fact the generic)
possibility is anti-deSitter (AdS) supersymmetry where $W|_{0}\ne0$
so that the cosmological constant is $V|_{0}=-3e^{K}|W|^{2}$. Perturbative
corrections around such a vacuum can in principle break supersymmetry.

Here we will consider only the leading, generically non-vanishing,
quadratic divergence that arises in perturbation theory. This (as
well as the log divergences and the finite terms to one loop) has
been calculated for an arbitrary curved background \cite{Barbieri:1983bv}\cite{Gaillard:1993es}
and gives the result\[
\delta L=\frac{\Lambda^{2}}{32\pi^{2}}[\frac{N+1}{2}r-2(N-5)V-2e^{G}\{(N-1)-H^{i\bar{j}}G_{i}G_{\bar{j}}\}].\]
Here $r$ is the Ricci scalar of space time and \begin{eqnarray}
H_{i\bar{j}}= & R_{i\bar{j}}+F_{i\bar{j}}\label{Hijbar}\\
R_{i\bar{j}}= & \partial_{i}\partial_{\bar{j}}\ln\det G_{m\bar{n}}\nonumber \\
F_{i\bar{j}}= & -\partial_{i}\partial_{\bar{j}}\ln\det Re[f_{ab}].\nonumber \end{eqnarray}

Here $f_{ab}(z,\bar{z})$ is the gauge coupling function. Thus keeping
only these quadratic divergences the one loop corrected Lagrangian
can then be written as \[
L=\frac{\tilde{M}_{p}^{2}}{2}r-\tilde{V}.\]
Here $\tilde{M}_{p}^{2}=(1+\frac{\Lambda^{2}}{32\pi^{2}}\frac{N+1}{2})$
is the corrected Plank mass and the corrected potential is \begin{equation}
\tilde{V}=e^{G}[G_{i}(K^{i\bar{j}}-\beta H^{i\bar{j}})G_{\bar{j}}-\{3-\beta(N-1)\}]\label{v1loop}\end{equation}
 with $G$ being redefined by adding a constant $\alpha=1+\frac{\Lambda^{2}}{16\pi^{2}}(N-5)$
and $\beta=\frac{\Lambda^{2}}{16\pi^{2}\alpha}$ . Indeed if the perturbative
corrections are computed with a regularization scheme which preserves
the (local) supersymmetry of the action, then one should be able to
write it in the same form as the classical potential, \begin{equation}
\tilde{V}=e^{\tilde{G}}(\tilde{G}_{i}\tilde{G}^{i\bar{j}}\tilde{G}_{\bar{j}}-3).\label{vtilde}\end{equation}
Here $\tilde{G}=G+\delta K$ where the second term is the pertubative
correction to the Kaehler potential. This can be computed from the
corrections to the kinetic terms of the chiral scalars of the theory.
In fact reference \cite{Gaillard:1993es} does give expressions for
the latter. However it is far from obvious how to express the potential
in the above form if one assumed that the calculations of \cite{Gaillard:1993es}
respected the SUGRA symmetry. So we will simply continue to work with
(\ref{v1loop}) since we are only interested in the qualitative and
order of magnitude effects of these corrections %
\footnote{It should be noted that (\ref{v1loop}) is in agreement with equations
(1.1) and (1.13) of \cite{Ferrara:1994kg} upto the constant shift
of $G$.%
}, but will assume that there is a correction to the Kaehler potential
that gurantees the validity of (\ref{vtilde}).

\section{Calculating MSSM parameters}

To calculate MSSM parameters one expands the Kaehler potential and
the superpotential in terms of the MSSM fields $\phi^{i}$,\begin{eqnarray}
K & = & \hat{K}+\tilde{K}_{i\bar{j}}\phi^{i}\bar{\phi}^{\bar{j}}+Z_{ij}\phi^{i}\phi^{j}+\ldots,\label{Kexpn}\\
W & = & \hat{W}+\mu_{ij}\phi^{i}\phi^{j}+Y_{ijk}\phi^{i}\phi^{j}\phi^{k}+\ldots.\label{Wexpn}\end{eqnarray}
 The coefficients of the powers of the MSSM fields are functions of
(the stabilized values of) the hidden (moduli) sector fields. Let
us go to a basis in which $\tilde{K}_{i\bar{j}}=\tilde{K}_{i}\delta_{ij}$.
The Yukawa couplings and the soft terms of the canonically normalized
fields are \cite{Kaplunovsky:1993rd,Brignole:1997dp}, \begin{eqnarray}
\hat{Y}_{ijk} & = & Y_{ijk}\frac{\bar{\hat{W}}}{|W|}e^{\hat{K}/2}(\tilde{K}_{i}\tilde{K}_{j}\tilde{K}_{k})^{-1/2}\label{Y}\\
A_{ijk} & = & F^{m}(\hat{K}_{m}+\partial_{m}(\ln Y_{ijk}-\ln(\tilde{K}_{i}\tilde{K}_{j}\tilde{K}_{k}))\label{A}\\
m_{i}^{2} & = & m_{3/2}^{2}+V_{0}-F^{m}\bar{F}^{\bar{n}}\partial_{m}\partial_{\bar{n}}\ln\tilde{K}_{i}\label{m}\\
M_{a} & = & \frac{1}{2}(\Re f_{a})^{-1}F^{m}\partial_{m}f_{a}\label{M}\end{eqnarray}
 These expressions are supposed to be valid at some high scale - presumably
somewhat below the string or KK scale. What is usually done is to
use these in a model where the classical CC has been fine-tuned to
be close to zero and positive by one or other mechanism, such as adding
the uplift terms of the original KKLT model or D terms. Then the physical
predictions are supposed to be obtained by running these down to the
MSSM scale by using the RG equations.

This procedure is however not meaningful. It is simply incorrect to
ignore the running of the cosmological constant, which generically
runs quadratically as was discussed in section \ref{sec:One-loop-effective}.
In fact it is the final CC (at the largest scales) which has a measured
value at the $10^{-3}eV$scale. The problem is that this final condition
has to be used to determine the initial conditions that are needed
to evolve the MSSM parameters!

To be concrete let us assume that a model with three generations and
the standard model gauge group has been found with the moduli stabilized
by a combination of fluxes and non-perturbative terms as in the KKLT
model. Let us denote the set of stringy parameters (data on the internal
manifold and the fluxes) by $ $$\beta$. This set determines the
moduli potential and hence in particular the cosmological constant
and the coefficients in (\ref{Kexpn})(\ref{Wexpn}). What is usually
done \cite{Grana:2003ek}\cite{Camara:2003ku}\cite{Lust:2004fi}\cite{Choi:2005ge}\cite{Allanach:2005pv}\cite{Conlon:2005ki}
following the original arguments of \cite{Kaplunovsky:1993rd,Brignole:1997dp}
is to tune the CC to be zero (or positive and small) by an appropriate
choice of $\beta$. However as we argued in section \ref{sec:One-loop-effective}
if one tuned the CC at this high scale to be small, one would get
a large CC at the MSSM scale (and below).

Thus a meaningful calculation of the MSSM parameters must have a starting
point, which at the classical string level has a CC which is tuned
such that the final CC, including all quantum corrections, is in accordance
with observation. This is of course the well-known non-trivial part
of the fine-tuning problem of the CC. In the context of classical
string solutions of the landscape this probem takes on the following
aspect. In terms of RG evolution the condition on the CC is a final
condition (looking at the flow as one from the UV to the IR). In other
words finding the set of allowed $\beta$ means solving an equation
of the form (as we discussed in section \ref{sec:Curved} the actual
condition is even more complicated)\begin{equation}
V_{eff,0}(\beta)=V_{c,0}(\beta)+\frac{1}{32\pi^{2}}{\rm Str}M^{2}(\beta)\Lambda^{2}|_{0}+\frac{1}{64\pi^{2}}{\rm Str}M^{4}(\beta)\ln\frac{M^{2}(\beta)}{\Lambda^{2}}|_{0}+\ldots.\simeq O((10^{-120})\label{CCobs}\end{equation}
where the last estimate is written in Planck units. Here $V_{c}$
denotes the classical potential, $V_{eff}$ denotes the quantum effective
potential and the subscript $0$ denotes evaluation at the minimum
of the effective potential. Fixing this final value within experimental
errors will lead to a large subspace of values of $\beta$ that is
not only (generically) going to be much larger than the subspace that
yields a classical CC that is within the bounds, but is also much
harder to determine. This is because the inverse RG problem depends
on knowing the supermultiplet splittings which of course are dependent
on the set $\beta$. Once this space of $\beta$ is determined the
initial value space for the RG evolution can be determined and then
one could calculate for each point in this space the corresponding
MSSM parameters at the relevant scale. Needless to say this appears
to be a rather intractable problem!

The point is that quantum corrections, should not be expected to cancel
amongst themselves barring some miracle. Let us try to estimate them.
From (\ref{CCobs}) we see that\begin{equation}
\delta V|_{0}\equiv\frac{1}{32\pi^{2}}{\rm Str}M^{2}\Lambda^{2}|_{0}+\frac{1}{64\pi^{2}}{\rm Str}M^{4}\ln\frac{M^{2}}{\Lambda^{2}}|_{0}+\ldots.=O(10^{-46}).\label{correctionest}\end{equation}
In the last step we have made an estimate of the typical value of
the radiative corrections after supersymmetry breaking by taking the
cutoff to be at an intermediate scale at $10^{-8}$ as in many popular
models (see for example \cite{Conlon:2005ki}) and estimating the
squared mass splitting in typical SUSY multiplets to be around a $TeV^{2}=10^{-30}$,
from the lower limit on SUSY partners. This means that in order to
get the observed value (barring some highly unlikely cancellation
of the quantum effects among themselves) one needs a classical value
$V_{c}|_{0}\sim O(10^{-46})$. Even the most conservative estimate
of this (for instance in those models where the quadratic divergence
is absent) gives a value $O(10^{-60})$ which is 60 orders of magnitude
larger than the value of the observed CC.

On the other hand one might ask whether the problem of the CC can
be decoupled from the calculation of the MSSM parameters. In other
words if we calculate the latter after tuning the CC at the classical
level to the observed value and ignore the quantum corrections to
the CC, can we get MSSM parameter values that are approximately correct.
This is what is typically assumed in the literature. However the point
is that the classical Kaehler potential $K$ AND superpotential $W$
for a theory tuned (i.e. $\beta$ chosen) such that (\ref{CCobs})
is satisfied, will in general be very different from that in a theory
in which the classical potential by itself is tuned to satisfy the
observational constraint. There is no perturbative relation between
the two sets of $K$'s and $W$'s. They are obtained in different
flux sectors and there is no reason at all to think that even qualitative
predictions obtained with the latter theory will be obtained also
in the former theory. 

$ $The essential point can be summarized as follows. Suppose the
tuning $V_{c}(\beta)|_{0}=O(10^{-120})$ yields a subspace ${\cal L}_{c}$
of the landscape ${\cal L}$ i.e. $\beta=\beta_{c}\in{\cal L}_{c}\subset{\cal L}$.
On the other hand the tuning $V_{eff,0}(\beta)=O(10^{-120})$ will
yield $\beta=\beta_{eff}\in{\cal L}_{eff}\subset{\cal L}$. Barring
accidental cancellations that would make the quantum corrections in
formulas such as (\ref{correctionest}) of the order of the observed
CC, one should expect that ${\cal L}_{eff}\cap{\cal L}_{c}=0$. There
is absolutely no reason to expect in general that $\tilde{K}_{eff,i}(\beta_{eff})$
and $W(\beta_{eff})$ with $\beta_{eff}\subset{\cal L}_{eff}$ are
approximately equal (respectively) to $\tilde{K}_{ci}(\beta_{c}),\, W(\beta_{c})$,
with $\beta_{c}\subset{\cal L}_{c}$. It should be emphasized here
that the values $\beta$, enter the formulae for the MSSM parameters
both directly, and indirectly through the dependence of the moduli
at the minimum on them. Thus the values of the measurable quantities
(\ref{Y},\ref{A},\ref{m},\ref{M}) in the quantum theory are likely
to be very different from the values obtained in the theory with the
cosmological constant tuned to zero in the classical theory. 

Actually even if one knew the set of $\beta_{eff}$ defined in the
previous paragraph it still does not make sense to compute the soft
terms with the classical value of the Kaehler potential. Thus consider
the effective quantum potential $V(K_{eff},W(\beta_{eff})$, where
$K_{eff}=K_{c}+\delta K$ (the second term includes both perturbative
and non-perturbative corrections). The superpotential $W$ is of course
not supposed to be renormalized in perturbation theory but contains
non-perturbative effects as in KKLT. The physical cosmological constant
is then \begin{equation}
V_{eff,0}\equiv V(K_{eff},W(\beta_{eff}))|_{0}=V(K_{c},W(\beta_{eff}))|_{0}+\delta K.\frac{\delta V}{\delta K}+O(\delta K^{2})\sim(10^{-3}eV)^{4}\label{eq:Keff}\end{equation}
$ $However as we've argued earlier the second term is generically
around 75 orders of magnitude larger than the observed CC and has
to be cancelled by a classical term which is of the same order. Thus
the corrections to the Kaehler potential have to be of the same order
as the classical Kahler potential and these corrections will give
corrections to the soft masses and couplings calculated from the classical
term that are not suppressed. In fact it is clear from the above analysis
what the basic (hidden) assumption that is made in typical string
phenomenology calculations is. It is that (at least for small fluctuations
around the relevant minimum)\begin{equation}
V(K_{c},W(\beta_{c}))\simeq V(K_{eff},W(\beta_{eff})).\label{eq:hidden}\end{equation}
Note that in the above not only is the Kaehler potential on the right
hand side the quantum corrected one, but also the superpotential is
evaluated at a different set of internal data. If indeed (\ref{eq:hidden})
is true, then the calculations made with the classical Kaehler potential
(tuned so that minium is the observed CC) will be in agreement with
that made with the effective potential tuned so that its minimum is
the observed CC). Unfortunately there is no a priori justification
for this assumption!

Given that there is no gurantee that (\ref{eq:hidden}) is correct
it is not clear that even qualitative predictions will survive. Except
under the very special circumstances discussed earlier, the only ones
that can be expected to survive are those that can be shown to be
independent of the flux parameters $\beta$ and the Kaehler potential.
Thus models in which certain Yukawa couplings vanish (irrespective
of the values of the fluxes) will retain this feature in the full
quantum theory upto (small?) non-perturbative corrections. Unfortunately
it is difficult to find any other classical predictions that can survive
the tuning of the CC in the effective quantum theory after supersymmetry
is broken.

\section{Conclusions}

The purpose of this note was to bring together certain well-known
facts which are central to understanding the relevance of string theory
to physics at the Large Hadron Collider (LHC). The most important
of these is the universally acknowledged but (in the context of KKLT
models) universally ignored fact that the CC undergoes quantum corrections
\footnote{For an early (pre-lanscape) attempt to incorporate this into the calculation
of MSSM parameters see \cite{Choi:1994xg,Choi:1997de}. But these
works do not take into acount the fact that any attempt to fine tune
the CC at the quantum effective level changes the fine tuning parameters
and thus can in general completely change both the Kaehler and the
superpotential from the classically tuned values.%
}. The BP argument \cite{Bousso:2000xa} of course tried to take the
latter into account, but as we pointed out this is not relevant to
what happens in concrete string theory situations exemplified by KKLT
type constructions. We have argued that since these are essentially
classical arguments (with the addition of some non-perturbative corrections
in the type IIB case) it is not meaningful to demand that the value
of the minimum of the potential be at the observed value of the CC.
Even if one is to make qualitative predictions or statistical ones
(as advocated by Douglas and collaborators \cite{Douglas:2004zg})
one has to face up to the fact that the quantum effects can completely
change any classical predictions, essentially because of the cosmological
constant problem. Clearly the Anthropic Principle is of no use in
resolving this issue.

Finally it should be noted that if a dynamical principle selected
one or a few vacua (perhaps after imposing some criteria such as the
dimensionality of space and the number of generations and the standard
model gauge group) then one can proceed in the top down fashion to
calculate for these models both the cosmological constant and the
MSSM parameters. This was of course the hope of string theorists until
very recently. The point of this note is merely to argue that if such
a principle does not exist, then it is unlikely that top-down approaches
have any relevance for phenomenology.

\section{Acknowledgments}

I wish to thank Ralph Blumenhagen, Oliver DeWolfe and especially Jan
Louis for discussions and comments on the manuscript. I also wish
to thank the latter for hospitality at DESY/ ITP-University of Hamburg
and Dieter Luest for hospitality at the Max-Planck Institute, Munich,
and the Aspen Center for Physics where this work was completed. In
addition I would like to thank Peter Nilles, Fernando Quevedo, Graham
Ross and other participants at the String Phenomenology workshop at
KITP for discussions on the revised version. I also wish to thank
the Perimeter Institute and DOE grant No. DE-FG02-91-ER-40672 for
partial support.

\bibliographystyle{apsrev} \bibliographystyle{apsrev}
\bibliography{myrefs}

\end{document}